
\magnification=\magstep2

\font\tenmibf=cmmib10
\textfont15=\tenmibf
\mathchardef\bepsilon="7F0F
\parskip\medskipamount

\hsize15.6truecm
\vsize22truecm
\hoffset1truecm
\voffset0.6truecm
\baselineskip0.7truecm
\tolerance1000
\overfullrule=0mm

\def\hE{{\displaystyle{\cal E }}}

\hfill{IASSNS-HEP-98/50}
\vskip0.6truecm
\centerline{\bf The Schr\"odinger-Like Equation for 
a Nonrelativistic Electron} 
\vskip0.1truecm
\centerline{\bf in a Photon Field of Arbitrary Intensity}
\vskip1.0truecm
\centerline{Dong-Sheng Guo$^{*}$ and R. R. Freeman}
\vskip0.1truecm
\centerline{Lawrence Livermore National Laboratory, 
Livermore, CA 94550}
\vskip0.4truecm
\centerline{Yong-Shi Wu$^{\dagger}$}
\vskip0.1truecm
\centerline{Institute for Advanced Study, Olden Lane, 
Princeton, NJ 08540} 
\vskip0.4truecm
\noindent({\bf Abstract})
The ordinary Schr\"odinger equation with minimal 
coupling for a nonrelativistic electron interacting 
with a single-mode photon field is not satisfied 
by the nonrelativistic limit of the exact 
solutions to the corresponding Dirac equation. A 
Schr\"odinger-like equation valid for arbitrary photon
intensity is derived from the Dirac equation without 
the weak-field assumption. The "eigenvalue" in the new 
equation is an operator in a Cartan subalgebra. An 
approximation consistent with the nonrelativistic 
energy level derived from its relativistic value 
replaces the "eigenvalue" operator by an ordinary 
number, recovering the ordinary Schr\"odinger eigenvalue 
equation used in the formal scattering formalism. The 
Schr\"odinger-like equation for the multimode case 
is also presented.
\vskip0.1truecm
\noindent-------------------------\hfill\break
\noindent $^*$ Permanent address:  Department of Physics, Southern 
University and A\&M College, Baton Rouge, LA 70813. \hfill\break
\noindent $^{\dagger}$ On sabbatical leave from Department of Physics, 
University of Utah, Salt Lake City, UT 84112. 

\filbreak
 
The exact analytical solutions for an electron interacting with a
photon field play a very important role in theories and
calculations of multiphoton ionization (MPI) and multiphoton
scattering processes [1-8]. These theoretical results have achieved
notable agreements [2,4,8] with experiments done during late 1980s
[9-11]. However, rigorously speaking, the theoretical treatments 
for a nonrelativistic (NR) electron interacting with a photon field [3-6] 
had a logical 
loophole; the purpose of this note is to fill this loophole. A
bonus of our efforts is the derivation of a Schr\"odinger-like
equation for an NR electron interacting with a photon field 
that is valid for arbitrary photon intensity (within the range
consistent with the NR motion of the electron).   

Before proceeding to formal considerations, let us first point out an
important feature of the MPI phenomenon. Namely, the total energy of
the photons interacting with the electron in a strong radiation field
can be comparable to the electron mass.  For example, the photon
energy of 1064 $nm$ is of the order 1 $eV$, when the laser intensity
is of the order $10^{13}$ $watt/cm^2$, the ponderomotive number is of
order of unity. This number is the photon number in a disc volume
$V_p$ with thickness as the electron classical radius $r_c$ and the
cross section made by the radius of the photon circular wavelength
$\lambda/2\pi$.  The interaction volume of an atomic electron $V$ can
be regarded as a disk volume with the thickness as the Bohr radius and
the same cross section of $V_p$. Thus, $V=137^2 V_p$. At the mentioned
intensity the background photon number is about $2\times 10^4$ with
total energy of the order $2\times 10^4$ $eV$. If one increases the
intensity of the light with the same wavelength to $2.5\times 10^{14}$
$watt/cm^2$, the total interacting background photon energy will be
around the electron mass. So the weak-field approximation used in the
usual quantum electrodynamics does not apply here. We are confronting
the problem to get a non-relativistic approximation with arbitrary
photon intensity.  The problem addressed in this paper is realistic
and in the range of current experiments.

The previous treatments [3-8] of the MPI theory adopt a
second-quantized formulation for the laser field. The equations of
motion, in the Heisenberg picture, for a relativistic
(first-quantized) electron interacting with a photon field is the
well-known Dirac equation which, say for the case of a single-mode
laser, reads
$$[i\gamma\partial-e\gamma A (k  x)-m_e]\Psi(x)=0,\eqno(1)$$
where 
$$A (k  x)=g (\epsilon a  e^{-ik  x}+\epsilon^* a^{\dag} e^{ik  x}), 
\eqno(2) $$
with $a$ and $a^{\dag}$, respectively, the photon annihilation and
creation operator, and $g =(2V_{\gamma} 
\omega)^{-1/2}$, $V_{\gamma}$ being the normalization volume of 
the photon field, and the polarization four-vector $\epsilon =
(0,\bepsilon)$. The Dirac equation has been solved exactly either with
a single-mode photon field [1] or with a multimode photon field which
propagates in one direction [7]. The NR limit of these exact solutions
has been derived in ref. [2], and one is tempted to use them in the the
theory for MPI in which the emitted electrons are nonrelativistic. 

As usual, the MPI theory started with the ordinary Schr\"odinger 
equation with the standard minimum coupling [12], which in 
the Schr\"odinger picture was the eigenvalue equation 
$$H \Psi({\bf r})={\hE } \Psi({\bf r}), \eqno(3)$$
with the Hamiltonian 
$$H={1\over 2m_e}[-i\nabla-e{\bf A}(-{\bf k}
\cdot {\bf r})]^2+\omega N_a ,
\eqno(4)$$
where 
$${\bf A}(-{\bf k}\cdot{\bf r})
=g( {\bepsilon} e^{i{\bf k}\cdot {\bf r}}a
+ {\bepsilon} ^*e^{-i{\bf k}\cdot {\bf r}}a^{\dag}) ,
\eqno(5)$$ 
and $g=(2V_{\gamma}\omega)^{-1/2}$, $V_{\gamma}$ being the
normalization volume of the photon field. $N_a$ is the photon 
number operator:
$$N_a={1\over 2} (aa^{\dag}+a^{\dag} a).\eqno(6)$$ 
The polarization vectors ${\bepsilon}$ and ${\bepsilon}^*$ 
are defined by
$${\bepsilon}=[{\bepsilon}_x \cos(\xi/2)+i{\bepsilon}_y 
\sin(\xi/2)]e^{i\Theta/2},$$
$${\bepsilon}^*=[{\bepsilon}_x \cos(\xi/2)-i{\bepsilon}_y 
\sin(\xi/2)]e^{-i\Theta/2}, \eqno(7)$$
and satisfy
$${\bepsilon}\cdot {\bepsilon}^*=1 ,\quad
{\bepsilon}\cdot {\bepsilon}=\cos\xi e^{i \Theta},\quad
{\bepsilon}^* \cdot {\bepsilon}^* =\cos\xi e^{-i \Theta}.$$
The angle $\xi$ determines the degree of polarization, such that $\xi
= \pi/2$ corresponds to circular polarization and $\xi = 0 $, to
linear polarization. (The phase angle $\Theta/2$ is introduced to
characterize the initial phase value of the photon mode in an earlier
work [3]. With this phase, a full "squeezed light" transformation [13]
can be fulfilled in the solution process. In multimode cases, the
relative value of this phase for each mode will be important.)

In the following, we first show that the Schr\"odinger eigenvalue
equation (3) with the NR Hamiltonian (4) is not satisfied by the NR
wavefunctions obtained from the exact solutions to the
corresponding Dirac equation (1). To see this, let us remove the
coordinate dependence of the ${\bf A}(-{\bf k}\cdot{\bf r})$ field by
applying a canonical transformation [14,15]
$$\Psi({\bf r})=e^{-i {\bf k}\cdot 
{\bf r}N_a} \phi({\bf r}). \eqno(8)$$
Equation (3) then becomes
$$\lbrace {1\over 2m_e}(-i\nabla-{\bf k}N_a)^2
- {e\over 2m_e}[(-i\nabla)\cdot {\bf A}+{\bf A}\cdot 
(-i\nabla)]$$
$$
+{e^2{\bf A}^2\over 2m_e} +\omega N_a \rbrace \phi({\bf r})
={\hE } \phi({\bf r}),
\eqno(9)$$  
where ${\bf k}\cdot{\bf A}=0$ by transversality. Here, ${\bf A}$ is 
coordinate-independent and defined as
$${\bf A}=e^{i {\bf k}\cdot {\bf r}N_a}{\bf A}(-{\bf k}\cdot{\bf r})
e^{-i {\bf k}\cdot {\bf r}N_a}=g ({\bepsilon} a +{\bepsilon}^* a^{\dag}) .
\eqno(10)$$

Setting $\phi ({\bf r})=e^{i{\bf p}\cdot {\bf r}}\phi$, we obtain
the coordinate-independent equation: 
$$[{1\over 2m_e}({\bf p}-{\bf k}N_a)^2
-{e\over m_e}{\bf p}\cdot{\bf A}
+{e^2{\bf A}^2\over 2m_e}+\omega N_a]\phi={\hE } \phi .
\eqno(11)$$  
Now we note that the term $({\bf k}N_a)^2\equiv{\bf k}N_a
\cdot{\bf k}N_a$ in Eqs. (9) and (11) does not exist in the
Dirac equation (1) and its squared form, which contain the 
creation and annihilation operators only up to quadratic terms. 
The exact solutions to the Dirac equation and their NR limit 
were obtained from the photon Fock states, i.e. the number states, 
by only "squeezed light" and "coherent light" transformations [1]. 
Any equation satisfied by these states can consist of operators 
$a$ or $a^{\dag}$ only up to quadratic ones.  Thus, the known NR
wavefunctions [16] do not satisfy the NR Schr\"odinger eigenvalue
equation (3).  

To justify the use of the NR wavefunctions, one of the authors has
introduced a special ansatz [3] which allows the replacement in
Eq. (11) of the ${\bf k} N_{a}$ terms by $\kappa {\bf k}$ with $\kappa$
a real number to be determined. The implicit assumption behind this
ansatz is that the corrections caused by this replacement should be at
most comparable to relativistic effects. With this ansatz and certain
covariance requirements, the solutions turned out to be just the NR
wavefunctions obtained by the NR limit from the exact solutions of the
Dirac equations.  Later, the ansatz was extended to the cases of
multimode photon fields with multiple propagation directions
[3,4,6]. Though this procedure leads to the correct NR wavefunctions 
in the single-mode case,
the reasons why the ansatz works were not explained. Moreover, the
validity of using the Schr\"odinger eigenvalue equation to describe an
NR electron in a strong photon field and the validity of the ansatz in  
multimode cases have never been rigorously justified.

Unlike the classical-field treatment, where the light field is treated
as an external field, our quantum field-theoretical approach for
photons requires a careful treatment to maintain relativistic
invariance for the photon field, while only the electron is to be
considered as an NR particle. The correct equation of motion should be
derived from the Dirac equation in the Schr\"odinger picture:
$$(H_e+H_\gamma+V)\Psi({\bf r})=p_0 \Psi({\bf r}),\eqno(12)$$
where
$$H_e={\bf \alpha}\cdot(-i\nabla)+\beta m_e , $$
$$H_\gamma=\omega N_a={\omega\over2}(aa^{\dag}+a^{\dag}a),\eqno(13)$$ 
$$V=-e {\bf\alpha}\cdot{\bf A}(-{\bf k}\cdot{\bf r}),$$ 
where
$\Psi({\bf r})=\left(\matrix{\Psi_1({\bf r})\cr
            \Psi_2({\bf r})\cr}\right),$ and 
$${\bf \alpha}=\left(\matrix{0&{\bf\sigma}\cr
            {\bf \sigma}&0\cr}\right),\quad\quad 
\beta=\left(\matrix{I&0\cr
            0&-I\cr}\right),\eqno(14)$$ where $\Psi_1({\bf r})$ and
$\Psi_2({\bf r})$ are the major and minor component respectively.
Thus Eq. (12) can be written as
$${\bf\sigma}\cdot[-i\nabla-e{\bf A}(-{\bf k}\cdot{\bf r})]
\Psi_2({\bf r})+ (m_e+\omega N_a)\Psi_1({\bf r})=p_0\Psi_1({\bf r}), $$
$${\bf\sigma}\cdot[-i\nabla-e{\bf A}(-{\bf k}\cdot{\bf r})]
\Psi_1({\bf r})+ (-m_e+\omega N_a)\Psi_2({\bf r})=p_0\Psi_2({\bf r}).
\eqno(15) $$
{}From the second equation of Eq. (15) we have
$$\Psi_2({\bf r})=(p_0+m_e-\omega N_a)^{-1}{\bf\sigma}\cdot
[-i\nabla-e{\bf A}(-{\bf k}\cdot{\bf r})]\Psi_1({\bf r}),\eqno(16)$$
Substituting $\Psi_2({\bf r})$ in the first equation of Eq. (15) and
ignoring the term ${\bf\sigma}\cdot[-e{\bf A}(-{\bf k}\cdot{\bf r})]
\Psi_2({\bf r})$ that pertains to the minor component, we obtain a
solo equation for the major component
$$\lbrace {\bf\sigma}\cdot[-i\nabla-e{\bf A}(-{\bf k}\cdot{\bf r})]
\rbrace^2\Psi_1({\bf r})=[(p_0-\omega N_a)^2-m_e^2]\Psi_1({\bf r}).
\eqno(17)$$
By neglecting the coupling between electron spin and photon
polarization, i.e. the term of $[{\bf\sigma}\cdot\bepsilon,
{\bf\sigma}\cdot\bepsilon^*]$, we have (writing $\Psi_1$ as $\Psi$)
$$[-i\nabla-e{\bf A}(-{\bf k}\cdot{\bf r})]^2
\Psi({\bf r})=[(p_0-\omega N_a)^2-m_e^2]\Psi({\bf r}).\eqno(18)$$
This equation can be written as an eigenvalue-like equation
 $$\{ {1\over2m_e}[-i\nabla-e{\bf A}(-{\bf k}\cdot{\bf r})]^2
+\omega N_a\}\Psi({\bf r})=\hE(N_a)\Psi({\bf r}),\eqno(19)$$
with 
$$\hE(N_a)\equiv{1\over2 m_e}[(p_0-\omega N_a)^2-m_e^2]+\omega N_a.
\eqno(19')$$

Equation (19) can be solved exactly. The following are the main steps
to obtain the solutions. The canonical transformation given by Eq. (7)
removes the coordinate dependence. Thus the equation becomes
$$(-i\nabla-e{\bf A}-{\bf k}N_a)^2\phi({\bf r})=
[(p_0-\omega N_a)^2-m_e^2] \phi({\bf r}) .\eqno(20)$$  
By setting $\phi({\bf r}) = e^{i{\bf p}\cdot {\bf r}}\phi$, 
we get the coordinate 
independent equation
$$[{\bf p}^2-2e{\bf p}\cdot{\bf A}+e^2{\bf A}^2+2(p_0\omega
-{\bf p}\cdot{\bf k})N_a]\phi=(p_0^2-m_e^2)\phi.\eqno(21)$$
A "squeezed light" transformation
$$ a=\cosh\chi c+\sinh\chi e^{-i\Theta} c^{\dag},$$
$$ a^{\dag} =\sinh\chi e^{i\Theta}c+\cosh\chi c^{\dag},
\eqno(22)$$
and a "coherent light" transformation 
$$\phi=D^{\dag} \mid n\rangle_c, \quad\quad
D=\exp(-\delta c^{\dag} +\delta^*c), \eqno(23)$$
$$\delta = eg{\bf p}\cdot{{\bepsilon}_c}^*
/[(p_0\omega-{\bf p}\cdot{\bf k}+e^2 g^2)^2 
-e^4g^4\cos^2\xi]^{1\over2}, $$ 
can be introduced to simplify the equation. Finally we have exact
solutions for the Schr\"odinger-like equation (19) or its
equivalent form (18)
$$\Psi({\bf r})=V_e^{-{1\over2}}\exp[i(-{\bf k}N_a+{\bf p})\cdot 
{\bf r}] D^{\dag} \mid n \rangle_c ,\eqno(24)$$
where
$$\mid n\rangle_c={c^{\dag n}\over \sqrt{n!}}\mid 0\rangle_c ,$$
$$ \mid 0\rangle _c=(\cosh\chi)^{-{1\over2}}
\sum_{s=0}^\infty(\tanh\chi)^s({(2s-1)!!\over (2s)!!})^{1\over2} 
e^{-is\Theta} \mid 2s\rangle,\eqno(25)$$
$$\chi = -{1\over2}\tanh^{-1}({e^2g^2\cos\xi\over p_0\omega- {\bf
p}\cdot{\bf k} + e^2g^2}) .$$ 
Here $\mid 2s \rangle$ is the Fock state
with $2s$ photons in the single-mode. 
The number $p_0$ is determined by 
the following algebraic equation
$$p_0^2 - m_e^2={\bf p}^2+2C(n+{1\over2})-2e^2 g^2({\bf p}\cdot 
{\bepsilon}_c)({\bf p}\cdot {\bepsilon}_c^*) C^{-1},$$
$$\eqno(26)$$
$$C\equiv [(p_0\omega-{\bf p}\cdot{\bf k}+e^2 g^2)^2 
-e^4g^4\cos^2\xi]^{1\over2}.$$

The solutions are also the
eigenfunction of the momentum operator:
$$(-i\nabla+i{\bf k}N_a)\Psi({\bf r})={\bf p}\Psi({\bf r}),\eqno(27)$$
which shows that ${\bf p}$ is the total momentum of this system. The
total momentum ${\bf p}$ has a unique decomposition on the electron
mass shell with light-like component in the ${\bf k}$ direction [1,5]:
$${\bf p}={\bf P}+\kappa{\bf k},$$
$$ p_0=m_e+{{\bf P}^2\over 2m_e}+\kappa\omega,\eqno(28)$$
$$\kappa ={C(n+{1\over 2})\over m_e\omega}
-{e^2g^2({\bf P}\cdot{\bepsilon}_c)({\bf P}\cdot {\bepsilon}^*_c)
\over m_e\omega C},$$
$$\quad\quad\quad\quad\quad\to (n+{1\over2} + u_p),\quad\hbox{(in the strong 
laser field case)}$$
with replacing $p_0\omega-{\bf p}\cdot{\bf k}$ by $m_e\omega$ in $C$. 
Here $u_p$ is the ponderomotive energy in units of photon energy.
With the help of Eq. (28), the solutions can be expressed as
$$\Psi({\bf r})=V_e^{-1/2}exp[i(-{\bf k}N_a+{\bf P}+\kappa{\bf
k})\cdot {\bf r}] D^{\dag} \mid n \rangle_c.\eqno(29)
$$ 
This result agrees with the known NR limit [2] of the exact solutions
to the Dirac equation (1), because of the following relation in the NR
limit:
$$p_0\omega-{\bf p}\cdot{\bf k}= (m_e+{{\bf P}^2\over 2m_e})\omega
-{\bf P}\cdot{\bf k}$$
$$\to m_e\omega.\eqno(30)$$
This provides us a consistency check for our equations (19) and (19$'$).

We emphasize that the Schr\"odinger-like equation (19) that we have
derived in the NR limit is not an ordinary eigenvalue equation, since
``eigenvalue'' (19$'$) is an operator (rather than a real number), 
which is a quadratic element in the commuting subalgebra generated by
$N_{a}$ in the enveloping algebra of $a$ and $a^{\dag}$. This 
subalgebra is also called a Cartan subalgebra. 

Though our Eq. (19) has the satisfying feature that the known NR
wavefunctions solve it exactly, it does not fit well the formal
scattering formalism which requires the wave functions to satisfy a
true eigenvalue equation. We propose to resolve this problem, by
numerizing the ``eigenvalue'' operator to its stationary values.  In
quantum mechanics, one can obtain the energy eigenvalues of a quantum
system by the variational method. Actually all the eigenvalues are
stationary values of the operator, not necessarily the minimum value
except for the ground state.  Here we do not need any variational
method since the wave functions are exactly known. In the following we
show that the stationary values of the operator $\hE(N_a)$ do give the
correct energy levels of the NR system. By setting
$${d \hE(N_a) \over d(\omega N_a)}=0,\eqno(31)$$ 
and treating $\hE(N_a)$ as a function of $\omega N_a$
we find that the stationary value, at $ \omega N_a =(p_0-m_e)I$ with I being 
the identity operator, is
$$\hE(N_a)\equiv(p_0-m_e)I+{1\over2m_e}
(p_0-m_e-\omega N_a )^2\to \hE I, \eqno(32)$$
with
$$\hE\equiv p_0-m_e= {{\bf P}^2\over2m_e}+\kappa\omega,
\eqno(32')$$
which is nothing but the energy level [2] for the interacting system
of the NR electron and the photon field without including the rest
mass of the electron.  The omission of the quadratic term in Eq. (32) is a NR 
limit process. We also observe that by replacing $N_a\omega$
in Eq. (19$'$) either by $\kappa\omega$ or by $(n+1/2)\omega$, $\hE$ has
the same value within the tolerance of the NR limit, showing the
stationary nature of the value (32). With the replacement of the
``eigenvalue'' operator by its stationary value (32') in the
eigenvalue-like equation (19), we get an effective eigenvalue
equation, which just recovers the ordinary Schr\"odinger eigenvalue
equation (3) with the minimal-coupling Hamiltonian (4). It is this
effective Schr\"odinger equation together with the NR wavefunctions
(24) that were used in the previous treatments [3-8]. So the physical
predictions obtained there remain valid.

In this way, we are led to the following procedure for treating
an NR electron in a photon field, which could be generalized to the
cases of multimode and multipotentials:\hfill\break
\indent (1) Solve the Schr\"odinger-like "eigenvalue" equation (19) 
to obtain the wavefunctions.\hfill\break
\indent (2) Obtain the stationary values of the "eigenvalue" operator 
as the energy levels.\hfill\break
\indent (3) Replace the operator "eigenvalue" by its stationary 
value to obtain an effective Schr\"odinger eigenvalue equation to 
be used in the formal scattering formalism.

In ending this article, we present the Schr\"odinger-like equation for
an NR electron in a two-mode photon field for future studies as follows
$$\{ {1\over2m_e}[-i\nabla-e{\bf A}_1(-{\bf k}_1\cdot{\bf r})
-e{\bf A}_2(-{\bf k}_2\cdot{\bf r})]^2$$
$$+\omega_1 N_1+\omega_2 N_2\}\Psi ({\bf r})=\hE(N_1,N_2)\Psi ({\bf r}),
\eqno(33)$$
$$\hE(N_1,N_2)\equiv{1\over2 m_e}[(p_0-\omega_1 N_1-\omega_2 N_2)^2-m_e^2]
+\omega_1 N_1+\omega_2 N_2 . $$
The coordinate independent equation to solve is 
$$[({\bf p}-e{\bf A})^2 + 2p(k_1N_1+k_2N_2) 
+ 2e({\bf k}_1N_1+{\bf k}_2N_2)\cdot{\bf A}$$
$$-2k_1k_2N_1N_2]\phi
=(p_0^2-m_e^2)\phi,\eqno(34)$$
where 
$pk_i\equiv(p_0\omega_i-{\bf p}\cdot{\bf k}_i), (i=1,2)$
and $k_1k_2\equiv(\omega_1\omega_2-{\bf k}_1\cdot{\bf k}_2).$

Compared with Eq. (21), we see that this equation contains a higher
order operator term $N_{1}N_{2}$. Searching for solutions to this
equation is one of our targets in the future research.

To summarize, in this paper we have addressed carefully the 
problem of the equations of motion for a nonrelativistic electron 
interacting with a single-mode photon field, which is valid for 
arbitrary photon intensity. We first showed that the usual 
Schr\"odinger eigenvalue equation is not solved by the 
NR limit of the wavefunctions that exactly solve
the corresponding Dirac equation. Then a Schr\"odinger-like equation 
is derived from the Dirac equation without using the weak-field 
assumption. Though the ``eigenvalue'' is an operator in a Cartan 
subalgebra involving the photon number operator, the new equation 
has a simpler structure compared to the usual eigenvalue equation. 
An effective Schr\"odinger equation with ordinary eigenvalues, good 
in the NR limit, is achieved by replacing the 
``eigenvalue'' operator by a number, which then can be used in the 
formal scattering theory. The Schr\"odinger-like equation for the 
multi-mode case is also presented. 

One of us, D.S.G., is supported in part by NSF Grant No. PHY-9603083.
Y.S.W. thanks the Institute for Theoretical Physics, University of
California at Santa Barbara for warm hospitality, where his part of
the work was begun, and the Japan Society for the Promotion of 
Science for a Fellowship, and the Institute for Solid State
Physics, University of Tokyo and Prof. Kohmoto for warm hospitality,
where the work was continued.  His work was supported in part by
the NSF through Grant No. PHY-9601277 and by the Monell Foundation.

\vskip0.5truecm

\centerline{\bf REFERENCES}

 1. D.-S. Guo and T. \AA berg, J. Phys. A: Math. Gen. {\bf 21}, 
4577 (1988).

 2. D.-S. Guo, T. \AA berg, and B. Crasemann, Phys. Rev. A {\bf 40}, 
4997 (1989). 

 3. D.-S. Guo, Phys. Rev. A {\bf 42}, 4302 (1990). 

 4. D.-S. Guo  and G. W. F. Drake, Phys. Rev. A {\bf 45}, 6622 
(1992).

 5. D.-S. Guo and G. W. F. Drake, J. Phys. A: Math. Gen. {\bf 25}, 3383 
(1992).

 6. D.-S. Guo and G. W. F. Drake, J. Phys. A: Math. Gen. {\bf 25}, 5377 
(1992).

 7. D.-S. Guo, J. Gao and A. Chu, Phys. Rev. A {\bf 54}, 1087 (1996).

 8. J. Gao and D.-S. Guo, Phys. Rev. A {\bf 47}, 5080 (1993).

 9. P. H. Bucksbaum, D. W. Schumacher and M. Bashkansky, 
Phys. Rev. Lett. {\bf 61}, 1182 (1988).

10. R. R. Freeman (Seminars) (1990).

11. R. R. Freeman and P. H. Bucksbaum, J. Phys B: At. Mol. 
Opt. Phys. {\bf 24}, 325 (1991).

12. J. J. Sakurai, {\it Advanced Quantum Mechanics}, 6th printing, 
Addison-Wesley (1977).

13. R. Loudon and P. L. Knight, J. Mod. Opt. {\bf 34}, 709 (1987).

14. T. D. Lee, F. E. Low, and D. Pines, Phys. Rev. {\bf 90}, 297 (1953). 

15. M. Girardeau, Phys. Flu. {\bf 4}, 279 (1960).

16. Those who are curious about the explicit form of the NR
wavefunctions are advised to take a look at Eq. (24) or (29), which
coincides with the NR limit of the exact solutions to the Dirac
equation (1).

\end

^Z^Z^Z^Z^Z^Z^Z^Z^Z^Z^Z^Z^Z^Z^Z^Z^Z^Z^Z^Z^Z^Z^Z^Z^Z^Z^Z^Z^Z^Z^Z^Z^Z^Z^Z^Z^Z^Z^Z^Z^Z^Z^Z^Z^Z^Z^Z^Z^Z^Z^Z^Z^Z^Z


\magnification=\magstep2

\font\tenmibf=cmmib10
\textfont15=\tenmibf
\mathchardef\bepsilon="7F0F
\parskip\medskipamount

\hsize15.6truecm
\vsize22truecm
\hoffset1truecm
\voffset0.6truecm
\baselineskip0.7truecm
\tolerance1000
\overfullrule=0mm

\def\hE{{\displaystyle{\cal E }}}

\centerline{\bf The Schr\"odinger-Like Equation for 
a Nonrelativistic Electron} 
\vskip0.1truecm
\centerline{\bf in a Photon Field of Arbitrary Intensity}
\vskip1.0truecm
\centerline{Dong-Sheng Guo$^{*}$ and R. R. Freeman}
\vskip0.1truecm
\centerline{Lawrence Livermore National Laboratory, 
Livermore, CA 94550}
\vskip0.4truecm
\centerline{Yong-Shi Wu$^{\dagger}$}
\vskip0.1truecm
\centerline{Institute for Advanced Study, Olden Lane, 
Princeton, NJ 08540} 
\vskip0.4truecm
\noindent({\bf Abstract})
The Schr\"odinger equation with minimal 
coupling for a nonrelativistic electron interacting 
with a single-mode photon field is not satisfied 
by the nonrelativistic limit of the exact 
solutions to the corresponding Dirac equation. A 
Schr\"odinger-like equation valid for arbitrary photon
intensity is derived from the Dirac equation without 
the weak-field assumption. The "eigenvalue" in the new 
equation is an operator in a Cartan subalgebra. An 
approximation consistent with the nonrelativistic energy level derived from 
its relativistic value replaces the "eigenvalue" operator by an ordinary 
number, recovering the Schr\"odinger eigenvalue 
equation used in the formal scattering formalism. The 
Schr\"odinger-like equation for the multimode case 
is also presented.
\vskip0.1truecm
\noindent-------------------------\hfill\break
\noindent $^*$ Permanent address:  Department of Physics, Southern 
University and A\&M College, Baton Rouge, LA 70813. \hfill\break
\noindent $^{\dagger}$ On sabbatical leave from Department of Physics, 
University of Utah, Salt Lake City, UT 84112. 

\filbreak
 
The exact analytical solutions for an electron interacting with a
photon field play a very important role in theories and
calculations of multiphoton ionization (MPI) and multiphoton
scattering processes [1-8]. These theoretical results have achieved
notable agreements [2,4,8] with experiments done during late 1980s
[9-11]. However, rigorously speaking, the theoretical treatments 
for a nonrelativistic (NR) electron interacting with a photon field [3-6] 
had a logical 
loophole; the purpose of this note is to fill this loophole. A
bonus of our efforts is the derivation of a Schr\"odinger-like
equation for an NR electron interacting with a photon field 
that is valid for arbitrary photon intensity (within the range
consistent with the NR motion of the electron).   

Before proceeding to formal considerations, let us first point out an
important feature of the MPI phenomenon. Namely, the total energy of
the photons interacting with the electron in a strong radiation field
can be comparable to the electron mass.  For example, the photon
energy of 1064 $nm$ is of the order 1 $eV$, when the laser intensity
is of the order $10^{13}$ $watt/cm^2$, the ponderomotive number is of
order of unity. This number is the photon number in a disc volume
$V_p$ with thickness as the electron classical radius $r_c$ and the
cross section made by the radius of the photon circular wavelength
$\lambda/2\pi$.  The interaction volume of an atomic electron $V$ can
be regarded as a disk volume with the thickness as the Bohr radius and
the same cross section of $V_p$. Thus, $V=137^2 V_p$. At the mentioned
intensity the background photon number is about $2\times 10^4$ with
total energy of the order $2\times 10^4$ $eV$. If one increases the
intensity of the light with the same wavelength to $2.5\times 10^{14}$
$watt/cm^2$, the total interacting background photon energy will be
around the electron mass. So the weak-field approximation used in the
usual quantum electrodynamics does not apply here. We are confronting
the problem to get a non-relativistic approximation with arbitrary
photon intensity.  The problem addressed in this paper is realistic
and in the range of current experiments.

The previous treatments [3-8] of the MPI theory adopt a
second-quantized formulation for the laser field. The equations of
motion, in the Heisenberg picture, for a relativistic
(first-quantized) electron interacting with a photon field is the
well-known Dirac equation which, say for the case of a single-mode
laser, reads
$$[i\gamma\partial-e\gamma A (k  x)-m_e]\Psi(x)=0,\eqno(1)$$
where 
$$A (k  x)=g (\epsilon a  e^{-ik  x}+\epsilon^* a^{\dag} e^{ik  x}), 
\eqno(2) $$
with $a$ and $a^{\dag}$, respectively, the photon annihilation and
creation operator, and $g =(2V_{\gamma} 
\omega)^{-1/2}$, $V_{\gamma}$ being the normalization volume of 
the photon field, and the polarization four-vector $\epsilon =
(0,\bepsilon)$. The Dirac equation has been solved exactly either with
a single-mode photon field [1] or with a multimode photon field which
propagates in one direction [7]. The NR limit of these exact solutions
has been derived in ref. [2], and one is tempted to use them in the the
theory for MPI in which the emitted electrons are nonrelativistic. 

As usual, the MPI theory started with the Schr\"odinger 
equation with the standard minimum coupling [12], which in 
the Schr\"odinger picture was the eigenvalue equation 
$$H \Psi({\bf r})={\hE } \Psi({\bf r}), \eqno(3)$$
with the Hamiltonian 
$$H={1\over 2m_e}[-i\nabla-e{\bf A}(-{\bf k}
\cdot {\bf r})]^2+\omega N_a ,
\eqno(4)$$
where 
$${\bf A}(-{\bf k}\cdot{\bf r})
=g( {\bepsilon} e^{i{\bf k}\cdot {\bf r}}a
+ {\bepsilon} ^*e^{-i{\bf k}\cdot {\bf r}}a^{\dag}) ,
\eqno(5)$$ 
and $g=(2V_{\gamma}\omega)^{-1/2}$, $V_{\gamma}$ being the
normalization volume of the photon field. $N_a$ is the photon 
number operator:
$$N_a={1\over 2} (aa^{\dag}+a^{\dag} a).\eqno(6)$$ 
The polarization vectors ${\bepsilon}$ and ${\bepsilon}^*$ 
are defined by
$${\bepsilon}=[{\bepsilon}_x \cos(\xi/2)+i{\bepsilon}_y 
\sin(\xi/2)]e^{i\Theta/2},$$
$${\bepsilon}^*=[{\bepsilon}_x \cos(\xi/2)-i{\bepsilon}_y 
\sin(\xi/2)]e^{-i\Theta/2}, \eqno(7)$$
and satisfy
$${\bepsilon}\cdot {\bepsilon}^*=1 ,\quad
{\bepsilon}\cdot {\bepsilon}=\cos\xi e^{i \Theta},\quad
{\bepsilon}^* \cdot {\bepsilon}^* =\cos\xi e^{-i \Theta}.$$
The angle $\xi$ determines the degree of polarization, such that $\xi
= \pi/2$ corresponds to circular polarization and $\xi = 0 $, to
linear polarization. (The phase angle $\Theta/2$ is introduced to
characterize the initial phase value of the photon mode in an earlier
work [3]. With this phase, a full "squeezed light" transformation [13]
can be fulfilled in the solution process. In multimode cases, the
relative value of this phase for each mode will be important.)

In the following, we first show that the Schr\"odinger eigenvalue
equation (3) with the NR Hamiltonian (4) is not satisfied by the NR
wavefunctions obtained from the exact solutions to the
corresponding Dirac equation (1). To see this, let us remove the
coordinate dependence of the ${\bf A}(-{\bf k}\cdot{\bf r})$ field by
applying a canonical transformation [14,15]
$$\Psi({\bf r})=e^{-i {\bf k}\cdot 
{\bf r}N_a} \phi({\bf r}). \eqno(8)$$
Equation (3) then becomes
$$\lbrace {1\over 2m_e}(-i\nabla-{\bf k}N_a)^2
- {e\over 2m_e}[(-i\nabla)\cdot {\bf A}+{\bf A}\cdot 
(-i\nabla)]$$
$$
+{e^2{\bf A}^2\over 2m_e} +\omega N_a \rbrace \phi({\bf r})
={\hE } \phi({\bf r}),
\eqno(9)$$  
where ${\bf k}\cdot{\bf A}=0$ by transversality. Here, ${\bf A}$ is 
coordinate-independent and defined as
$${\bf A}=e^{i {\bf k}\cdot {\bf r}N_a}{\bf A}(-{\bf k}\cdot{\bf r})
e^{-i {\bf k}\cdot {\bf r}N_a}=g ({\bepsilon} a +{\bepsilon}^* a^{\dag}) .
\eqno(10)$$

Setting $\phi ({\bf r})=e^{i{\bf p}\cdot {\bf r}}\phi$, we obtain
the coordinate-independent equation: 
$$[{1\over 2m_e}({\bf p}-{\bf k}N_a)^2
-{e\over m_e}{\bf p}\cdot{\bf A}
+{e^2{\bf A}^2\over 2m_e}+\omega N_a]\phi={\hE } \phi .
\eqno(11)$$  
Now we note that the term $({\bf k}N_a)^2\equiv{\bf k}N_a
\cdot{\bf k}N_a$ in Eqs. (9) and (11) does not exist in the
Dirac equation (1) and its squared form, which contain the 
creation and annihilation operators only up to quadratic terms. 
The exact solutions to the Dirac equation and their NR limit 
were obtained from the photon Fock states, i.e. the number states, 
by only "squeezed light" and "coherent light" transformations [1]. Any
equation satisfied by these states can consist of operators $a$ or
$a^{\dag}$ only up to quadratic ones.  Thus, the known NR
wavefunctions [16] do not satisfy the NR Schr\"odinger eigenvalue
equation (3).  

To justify the use of the NR wavefunctions, one of the authors has
introduced a special ansatz [3] which allows the replacement in
Eq. (11) of the ${\bf k} N_{a}$ terms by $\kappa {\bf k}$ with $\kappa$
a real number to be determined. The implicit assumption behind this
ansatz is that the corrections caused by this replacement should be at
most comparable to relativistic effects. With this ansatz and certain
covariance requirements, the solutions turned out to be just the NR
wavefunctions obtained by the NR limit from the exact solutions of the
Dirac equations.  Later, the ansatz was extended to the cases of
multimode photon fields with multiple propagation directions
[3,4,6]. Though this procedure leads to the correct NR wavefunctions 
in the single-mode case,
the reasons why the ansatz works were not explained. Moreover, the
validity of using the Schr\"odinger eigenvalue equation to describe an
NR electron in a strong photon field and the validity of the ansatz in  
multimode cases have never been rigorously justified.

Unlike the classical-field treatment, where the light field is treated
as an external field, our quantum field-theoretical approach for
photons requires a careful treatment to maintain relativistic
invariance for the photon field, while only the electron is to be
considered as an NR particle. The correct equation of motion should be
derived from the Dirac equation in the Schr\"odinger picture:
$$(H_e+H_\gamma+V)\Psi({\bf r})=p_0 \Psi({\bf r}),\eqno(12)$$
where
$$H_e={\bf \alpha}\cdot(-i\nabla)+\beta m_e , $$
$$H_\gamma=\omega N_a={\omega\over2}(aa^{\dag}+a^{\dag}a),\eqno(13)$$ 
$$V=-e {\bf\alpha}\cdot{\bf A}(-{\bf k}\cdot{\bf r}),$$ 
where
$\Psi({\bf r})=\left(\matrix{\Psi_1({\bf r})\cr
            \Psi_2({\bf r})\cr}\right),$ and 
$${\bf \alpha}=\left(\matrix{0&{\bf\sigma}\cr
            {\bf \sigma}&0\cr}\right),\quad\quad 
\beta=\left(\matrix{I&0\cr
            0&-I\cr}\right),\eqno(14)$$ where $\Psi_1({\bf r})$ and
$\Psi_2({\bf r})$ are the major and minor component respectively.
Thus Eq. (12) can be written as
$${\bf\sigma}\cdot[-i\nabla-e{\bf A}(-{\bf k}\cdot{\bf r})]
\Psi_2({\bf r})+ (m_e+\omega N_a)\Psi_1({\bf r})=p_0\Psi_1({\bf r}), $$
$${\bf\sigma}\cdot[-i\nabla-e{\bf A}(-{\bf k}\cdot{\bf r})]
\Psi_1({\bf r})+ (-m_e+\omega N_a)\Psi_2({\bf r})=p_0\Psi_2({\bf r}).
\eqno(15) $$
{}From the second equation of Eq. (15) we have
$$\Psi_2({\bf r})=(p_0+m_e-\omega N_a)^{-1}{\bf\sigma}\cdot
[-i\nabla-e{\bf A}(-{\bf k}\cdot{\bf r})]\Psi_1({\bf r}),\eqno(16)$$
Substituting $\Psi_2({\bf r})$ in the first equation of Eq. (15) and
ignoring the term ${\bf\sigma}\cdot[-e{\bf A}(-{\bf k}\cdot{\bf r})]
\Psi_2({\bf r})$ that pertains to the minor component, we obtain a
solo equation for the major component
$$\lbrace {\bf\sigma}\cdot[-i\nabla-e{\bf A}(-{\bf k}\cdot{\bf r})]
\rbrace^2\Psi_1({\bf r})=[(p_0-\omega N_a)^2-m_e^2]\Psi_1({\bf r}).
\eqno(17)$$
By neglecting the coupling between electron spin and photon
polarization, i.e. the term of $[{\bf\sigma}\cdot\bepsilon,
{\bf\sigma}\cdot\bepsilon^*]$, we have (writing $\Psi_1$ as $\Psi$)
$$[-i\nabla-e{\bf A}(-{\bf k}\cdot{\bf r})]^2
\Psi({\bf r})=[(p_0-\omega N_a)^2-m_e^2]\Psi({\bf r}).\eqno(18)$$
This equation can be written as an eigenvalue-like equation
 $$\{ {1\over2m_e}[-i\nabla-e{\bf A}(-{\bf k}\cdot{\bf r})]^2
+\omega N_a\}\Psi({\bf r})=\hE(N_a)\Psi({\bf r}),\eqno(19)$$
with 
$$\hE(N_a)\equiv{1\over2 m_e}[(p_0-\omega N_a)^2-m_e^2]+\omega N_a.
\eqno(19')$$

Equation (19) can be solved exactly. The following are the main steps
to obtain the solutions. The canonical transformation given by Eq. (8)
removes the coordinate dependence. Thus the equation becomes
$$(-i\nabla-e{\bf A}-{\bf k}N_a)^2\phi({\bf r})=
[(p_0-\omega N_a)^2-m_e^2] \phi({\bf r}) .\eqno(20)$$  
By setting $\phi({\bf r}) = e^{i{\bf p}\cdot {\bf r}}\phi$, 
we get the coordinate 
independent equation
$$[{\bf p}^2-2e{\bf p}\cdot{\bf A}+e^2{\bf A}^2+2(p_0\omega
-{\bf p}\cdot{\bf k})N_a]\phi=(p_0^2-m_e^2)\phi.\eqno(21)$$
A "squeezed light" transformation
$$ a=\cosh\chi c+\sinh\chi e^{-i\Theta} c^{\dag},$$
$$ a^{\dag} =\sinh\chi e^{i\Theta}c+\cosh\chi c^{\dag},
\eqno(22)$$
and a "coherent light" transformation 
$$\phi=D^{\dag} \mid n\rangle_c, \quad\quad
D=\exp(-\delta c^{\dag} +\delta^*c), \eqno(23)$$
$$\delta = eg{\bf p}\cdot{{\bepsilon}_c}^*
/[(p_0\omega-{\bf p}\cdot{\bf k}+e^2 g^2)^2 
-e^4g^4\cos^2\xi]^{1\over2}, $$ 
can be introduced to simplify the equation. Finally we have exact
solutions for the Schr\"odinger-like equation (19) or its
equivalent form (18)
$$\Psi({\bf r})=V_e^{-{1\over2}}\exp[i(-{\bf k}N_a+{\bf p})\cdot 
{\bf r}] D^{\dag} \mid n \rangle_c ,\eqno(24)$$
where
$$\mid n\rangle_c={c^{\dag n}\over \sqrt{n!}}\mid 0\rangle_c ,$$
$$ \mid 0\rangle _c=(\cosh\chi)^{-{1\over2}}
\sum_{s=0}^\infty(\tanh\chi)^s({(2s-1)!!\over (2s)!!})^{1\over2} 
e^{-is\Theta} \mid 2s\rangle,\eqno(25)$$
$$\chi = -{1\over2}\tanh^{-1}({e^2g^2\cos\xi\over p_0\omega- {\bf
p}\cdot{\bf k} + e^2g^2}) .$$ 
Here $\mid 2s \rangle$ is the Fock state
with $2s$ photons in the single-mode. 
The number $p_0$ is determined by 
the following algebraic equation
$$p_0^2 - m_e^2={\bf p}^2+2C(n+{1\over2})-2e^2 g^2({\bf p}\cdot 
{\bepsilon}_c)({\bf p}\cdot {\bepsilon}_c^*) C^{-1},$$
$$\eqno(26)$$
$$C\equiv [(p_0\omega-{\bf p}\cdot{\bf k}+e^2 g^2)^2 
-e^4g^4\cos^2\xi]^{1\over2}.$$

The solutions are also the
eigenfunction of the momentum operator:
$$(-i\nabla+i{\bf k}N_a)\Psi({\bf r})={\bf p}\Psi({\bf r}),\eqno(27)$$
which shows that ${\bf p}$ is the total momentum of this system. The
total momentum ${\bf p}$ has a unique decomposition on the electron
mass shell with light-like component in the ${\bf k}$ direction [1,5]:
$${\bf p}={\bf P}+\kappa{\bf k},$$
$$ p_0=m_e+{{\bf P}^2\over 2m_e}+\kappa\omega,\eqno(28)$$
$$\kappa ={C(n+{1\over 2})\over m_e\omega}
-{e^2g^2({\bf P}\cdot{\bepsilon}_c)({\bf P}\cdot {\bepsilon}^*_c)
\over m_e\omega C},$$
$$\quad\quad\quad\quad\quad\to (n+{1\over2} + u_p),\quad\hbox{(in the strong 
laser field case)}$$
with replacing $p_0\omega-{\bf p}\cdot{\bf k}$ by $m_e\omega$ in $C$. 
Here $u_p$ is the ponderomotive energy in units of photon energy.
With the help of Eq. (28), the solutions can be expressed as
$$\Psi({\bf r})=V_e^{-1/2}\exp[i(-{\bf k}N_a+{\bf P}+\kappa{\bf
k})\cdot {\bf r}] D^{\dag} \mid n \rangle_c.\eqno(29)
$$ 
This result agrees with the known NR limit [2] of the exact solutions
to the Dirac equation (1), because of the following relation in the NR
limit:
$$p_0\omega-{\bf p}\cdot{\bf k}= (m_e+{{\bf P}^2\over 2m_e})\omega
-{\bf P}\cdot{\bf k}$$
$$\to m_e\omega.\eqno(30)$$
This provides us a consistency check for our equations (19) and (19$'$).

We emphasize that the Schr\"odinger-like equation (19) that we have
derived in the NR limit is not an ordinary eigenvalue equation, since
``eigenvalue'' (19$'$) is an operator (rather than a real number), which
is a quadratic element in the commuting subalgebra generated by
$N_{a}$ in the enveloping algebra of $a$ and $a^{\dag}$. 

Though our Eq. (19) has the satisfying feature that the known NR
wavefunctions solve it exactly, it does not fit well the formal
scattering formalism which requires the wave functions to satisfy a
true eigenvalue equation. We propose to resolve this problem, by
numerizing the ``eigenvalue'' operator to its stationary values.  In
quantum mechanics, one can obtain the energy eigenvalues of a quantum
system by the variational method. Actually all the eigenvalues are
stationary values of the operator, not necessarily the minimum value
except for the ground state.  Here we do not need any variational
method since the wave functions are exactly known. In the following we
show that the stationary values of the operator $\hE(N_a)$ do give the
correct energy levels of the NR system. By setting
$${d \hE(N_a) \over d(\omega N_a)}=0,\eqno(31)$$ 
and treating $\hE(N_a)$ as a function of $\omega N_a$
we find that the stationary value, at $ \omega N_a =(p_0-m_e)I$ with $I$ 
being the identity operator, is
$$\hE(N_a)\equiv(p_0-m_e)I+{1\over2m_e}
(p_0-m_e-\omega N_a )^2\to \hE I, \eqno(32)$$
with
$$\hE\equiv p_0-m_e= {{\bf P}^2\over2m_e}+\kappa\omega,
\eqno(32')$$
which is nothing but the energy level [2] for the interacting system
of the NR electron and the photon field without including the rest
mass of the electron.  The omission of the quadratic term in Eq. (32) is a NR 
limit process. We also observe that by replacing $N_a\omega$
in Eq. (19$'$) either by $\kappa\omega$ or by $(n+1/2)\omega$, $\hE$ has
the same value within the tolerance of the NR limit, showing the
stationary nature of the value (32). With the replacement of the
``eigenvalue'' operator by its stationary value (32') in the
eigenvalue-like equation (19), we get an effective eigenvalue
equation, which just recovers the Schr\"odinger eigenvalue
equation (3) with the minimal-coupling Hamiltonian (4). It is this
effective Schr\"odinger equation together with the NR wavefunctions
(24) that were used in the previous treatments [3-8]. So the physical
predictions obtained there remain valid.

In this way, we are led to the following procedure for treating
an NR electron in a photon field, which could be generalized to the
cases of multimode and multipotentials:\hfill\break
\indent (1) Solve the Schr\"odinger-like "eigenvalue" equation (19) 
to obtain the wavefunctions.\hfill\break
\indent (2) Obtain the stationary values of the "eigenvalue" operator 
as the energy levels.\hfill\break
\indent (3) Replace the operator "eigenvalue" by its stationary 
value to obtain an effective Schr\"odinger eigenvalue equation to 
be used in the formal scattering formalism.

In ending this article, we present the Schr\"odinger-like equation for
an NR electron in a two-mode photon field for future studies as follows
$$\{ {1\over2m_e}[-i\nabla-e{\bf A}_1(-{\bf k}_1\cdot{\bf r})
-e{\bf A}_2(-{\bf k}_2\cdot{\bf r})]^2$$
$$+\omega_1 N_1+\omega_2 N_2\}\Psi ({\bf r})=\hE(N_1,N_2)\Psi ({\bf r}),
\eqno(33)$$
$$\hE(N_1,N_2)\equiv{1\over2 m_e}[(p_0-\omega_1 N_1-\omega_2 N_2)^2-m_e^2]
+\omega_1 N_1+\omega_2 N_2 . $$
The coordinate independent equation to solve is 
$$[({\bf p}-e{\bf A})^2 + 2p(k_1N_1+k_2N_2) 
+ 2e({\bf k}_1N_1+{\bf k}_2N_2)\cdot{\bf A}$$
$$-2k_1k_2N_1N_2]\phi
=(p_0^2-m_e^2)\phi,\eqno(34)$$
where 
$pk_i\equiv(p_0\omega_i-{\bf p}\cdot{\bf k}_i), (i=1,2)$
and $k_1k_2\equiv(\omega_1\omega_2-{\bf k}_1\cdot{\bf k}_2).$

Compared with Eq. (21), we see that this equation contains a higher
order operator term $N_{1}N_{2}$. Searching for solutions to this
equation is one of our targets in the future research.

To summarize, in this paper we have addressed carefully the 
problem of the equations of motion for a nonrelativistic electron 
interacting with a single-mode photon field, which is valid for 
arbitrary photon intensity. We first showed that the usual 
Schr\"odinger eigenvalue equation is not solved by the 
NR limit of the wavefunctions that exactly solve
the corresponding Dirac equation. Then a Schr\"odinger-like equation 
is derived from the Dirac equation without using the weak-field 
assumption. Though the ``eigenvalue'' is an operator in a Cartan 
subalgebra involving the photon number operator, the new equation 
has a simpler structure compared to the usual eigenvalue equation. 
An effective Schr\"odinger equation with ordinary eigenvalues, good 
in the NR limit, is achieved by replacing the 
``eigenvalue'' operator by a number, which then can be used in the 
formal scattering theory. The Schr\"odinger-like equation for the 
multi-mode case is also presented. 

One of us, D.S.G., is supported in part by NSF Grant No. PHY-9603083.
Y.S.W. thanks the Institute for Theoretical Physics, University of
California at Santa Barbara for warm hospitality, where his part of
the work was begun, and the Japan Society for the Promotion of 
Science for a Fellowship and the Institute for Solid State
Physics, University of Tokyo, and Prof. Kohmoto for warm hospitality,
where the work was continued.  His work was supported in part by
the NSF through Grant No. PHY-9601277 and by the Monell Foundation.

\vskip0.5truecm

\centerline{\bf REFERENCES}

 1. D.-S. Guo and T. \AA berg, J. Phys. A: Math. Gen. {\bf 21}, 
4577 (1988).

 2. D.-S. Guo, T. \AA berg, and B. Crasemann, Phys. Rev. A {\bf 40}, 
4997 (1989). 

 3. D.-S. Guo, Phys. Rev. A {\bf 42}, 4302 (1990). 

 4. D.-S. Guo  and G. W. F. Drake, Phys. Rev. A {\bf 45}, 6622 
(1992).

 5. D.-S. Guo and G. W. F. Drake, J. Phys. A: Math. Gen. {\bf 25}, 3383 
(1992).

 6. D.-S. Guo and G. W. F. Drake, J. Phys. A: Math. Gen. {\bf 25}, 5377 
(1992).

 7. D.-S. Guo, J. Gao and A. Chu, Phys. Rev. A {\bf 54}, 1087 (1996).

 8. J. Gao and D.-S. Guo, Phys. Rev. A {\bf 47}, 5080 (1993).

 9. P. H. Bucksbaum, D. W. Schumacher and M. Bashkansky, 
Phys. Rev. Lett. {\bf 61}, 1182 (1988).

10. R. R. Freeman (Seminars) (1990).

11. R. R. Freeman and P. H. Bucksbaum, J. Phys B: At. Mol. 
Opt. Phys. {\bf 24}, 325 (1991).

12. J. J. Sakurai, {\it Advanced Quantum Mechanics}, 6th printing, 
Addison-Wesley (1977).

13. R. Loudon and P. L. Knight, J. Mod. Opt. {\bf 34}, 709 (1987).

14. T. D. Lee, F. E. Low, and D. Pines, Phys. Rev. {\bf 90}, 297 (1953). 

15. M. Girardeau, Phys. Flu. {\bf 4}, 279 (1960).

16. Those who are curious about the explicit form of the NR
wavefunctions are advised to take a look at Eq. (24) or (29), which
coincides with the NR limit of the exact solutions to the Dirac
equation (1).

\end